\documentclass{iaus}
\usepackage{graphicx}
\def\apj{ApJ}

\def\apjl{ApJL}
\def\araa{ARA\&A}

\def\aap{A\&A}
\def\mnras{\textit{MNRAS}}

\def\lsim{~\raise0.3ex\hbox{$<$}\kern-0.75em{\lower0.65ex\hbox{$\sim$}}~}
\def\gsim{~\raise0.3ex\hbox{$>$}\kern-0.75em{\lower0.65ex\hbox{$\sim$}}~}

\title[ISM Turbulence] 
{Interstellar Turbulence and Star Formation}

\author[Kritsuk, Ustyugov \& Norman]   
{Alexei G. Kritsuk,$^1$ Sergey D. Ustyugov,$^2$
 \and Michael L. Norman$^1$}

\affiliation{$^1$UC San Diego, 9500 Gilman Drive MC 0424, La Jolla CA 92093-0424, USA \\ 
email: {\tt akritsuk@ucsd.edu, mlnorman@ucsd.edu} \\[\affilskip]
$^2${Keldysh Institute of Applied Mathematics, Miusskaya Sq.~4, 125047, Moscow, Russia
 \\email: {\tt ustyugs@keldysh.ru}}}

\pubyear{2010}
\volume{270}  
\pagerange{119--126}
\date{2010 July 25 and in revised form 2010 August X}
\jname{Computational Star Formation}
\editors{J. Alves, B. Elmegreen, J. M. Girart, \& V. Trimble, eds.}

\begin{document}

\maketitle

\begin{abstract}
We provide a brief overview of recent advances and outstanding 
issues in simulations of interstellar turbulence, including isothermal 
models for interior structure of molecular clouds and larger-scale 
multiphase models designed to simulate the formation of molecular 
clouds. We show how self-organization in highly compressible 
magnetized turbulence in the multiphase ISM can be exploited in
simple numerical models to generate realistic initial conditions
for star formation.

\end{abstract}

\firstsection 
\section{Introduction}
Since most of the ISM is characterized by very large Reynolds numbers, turbulent
motions control the structure of nearly all temperature and density 
regimes in the interstellar gas \cite[(Elmegreen \& Scalo 2004)]{elmegreen.04}.
Because of that, turbulence is often viewed as an organizing agent forming and shaping 
hierarchical cloudy structures in the diffuse ISM and ultimately in star-forming 
molecular clouds \cite[(e.g., V{\'a}zquez-Semadeni \& Passot 1999)]{vazquezsemadeni.99}.
Nonlinear advection dominating the dynamics of such highly compressible magnetized
multicsale self-gravitating flows makes computer simulations practically the only tool 
to study fundamental aspects of interstellar turbulence, even though effective Reynolds numbers 
in numerical models are always limited by the available computational resource 
\cite[(e.g., Kritsuk et al. 2006)]{kritsuk..06}.

Over the last five years three-dimensional numerical simulations fostered the 
development of theoretical concepts concerning the interstellar medium undergoing 
nonlinear self-interaction and self-organization in galactic disks. One can conventionally 
divide these models designed to tackle various aspects of interstellar turbulence into 
three different classes depending on the range of resolved scales and physics 
included: 
(i) {\em mesoscale} models that cover evolution of multiphase ISM in volumes with 
linear size of a few-to-ten kpc and resolve the flow structure down to a fraction of 1~pc 
\cite[(e.g., galactic fountain models developed by de Avilles \& Breitschwerdt 
2005-07)]{deavillez.05,deavillez.07};
(ii) {\em sub-mesoscale} models resolving the scale-height of the diffuse H{\sc i} ($\sim100$~pc) 
and usually limited to only warm-to-cold neutral phases (WNM and CNM) 
\cite[(e.g., Kissmann et al. 2008; Gazol et al. 2009; Gazol \& Kim 2010; Seifried et al. 
2010)]{kissmann...08,gazol..09,gazol.10,seifried..10}; and
(iii) {\em microscale} models for molecular cloud (MC) turbulence that assume an isothermal
equation of state and deal with $<10$~pc-sized subvolumes within MCs. {\em Global} galactic disk models 
\cite[(e.g., Tasker \& Bryan 2006-08, Wada 2008 and references therein)]{tasker.06,tasker.08,wada08} 
which represent the future of direct ISM turbulence modeling, are currently resolving
scales down to $\sim10$~pc, i.e. insufficient to properly follow the thermal structure 
of self-gravitating multiphase ISM.

Mesoscale models of supernova-powered (SNe) galactic fountain have demonstrated the
important role of dynamic pressure in the ISM that keeps large fractions of the gas mass
out of thermal equilibrium and elevates gas pressures of GMCs to the observed levels even
without direct action of self-gravity 
\cite[(Korpi et al. 1999; Mac Low et al. 2005; de Avilles \& Breitschwerdt 2005-07; Joung et al. 2006-09)]%
{korpi....99,maclow...05,deavillez.05,deavillez.07,joung.06,joung..09}. 
They also show that the effective integral scale of the SN-driven turbulence ($\sim75$~pc)
is about half the scale height of the H{\sc i} gas in the inner Galaxy 
[$100-150$~pc \cite[(Malhotra 1995)]{malhotra95}] and outline a general picture of 
probability distributions for the mass density, magnetic field strength, and thermal 
pressure in the turbulent ISM in disk-like galaxies.

Supersonic isothermal turbulence simulations in periodic boxes representative of the microscale
models provided many important insights into the physics of interstellar turbulence and helped 
to guide the interpretation of observations. These numerical experiments highlighted the importance 
of nonlinear advection as a major feature of compressible turbulence 
\cite[(Pouquet et al. 1991)]{pouquet..91}. At high Mach numbers, turbulent 
flows are dominated by shocks; therefore the velocity spectra are steeper than the Kolmogorov slope 
of $-5/3$ and closely resemble the Burgers $-2$ scaling \cite[(Kritsuk et al. 2006b)]{kritsuk...06}.
The physics of three-dimensional supersonic turbulence is, however, quite different from burgulence 
\cite[(Frisch \& Bec 2001)]{frisch.01} as the solenoidal velocity component always remains 
dominant \cite[(Pouquet et al. 1991; Pavlovski et al. 2006; Kritsuk et al. 2007; Pan et al. 2009; Schmidt et al. 2009; 
Kritsuk et al. 2010)]{pouquet..91,pavlovski..06,pan..09,schmidt....09,kritsuk...07,kritsuk...10}. At sonic 
Mach numbers $M_{\rm s}>3$, strong shock interactions and associated nonlinear instabilities create 
sophisticated multiscale pattern of nested {\sf U}-shaped structures in dynamically active 
regions morphologically similar to what is observed in molecular clouds
\cite[(Kritsuk et al. 2006a)]{kritsuk..06}. Scaling of the first-order velocity structure functions 
$S_1(\delta{\bf u})\sim\ell^{0.54}$ (where $\delta{\bf u}(\ell)={\bf u(x)-u(x+\hat{e}}\ell)$,
$S_p(\delta{\bf u})=\left<\left[\delta{\bf u}(\ell)\right]^p\right>$ and $\left<\ldots\right>$ indicates
averaging over an ensemble of random point pairs separated by the lag $\ell$) obtained in simulations 
\cite[(Kritsuk et al. 2007)]{kritsuk...07} is similar 
to the velocity scaling observed in molecular clouds $S_1(\delta{\bf u})\sim\ell^{0.56}$ 
\cite[(Heyer \& Brunt 2004)]{heyer.04}. Simulations also support the concept of (lossy) energy cascade 
in compressible turbulence \cite[(e.g., V{\'a}zquez-Semadeni et al. 2003)]{vazquezsemadeni..03},
suggesting that the kinetic energy directly lost in shocks constitutes a small fraction of the total 
energy dissipation. The fact that the Richardson-Kolmogorov cascade picture does approximately hold 
for supersonic turbulence follows from the linear scaling of the third-order structure function of
the mass-weighted velocity, $S_3(\delta\sqrt[3]{\rho}{\bf u})\sim\ell$, indicating constant turbulent 
energy transfer rate across the hierarchy of scales \cite[(Kritsuk et al. 2007; 
Kowal \& Lazarian 2007; Schwarz et al. 2010)]{kritsuk...07,kowal.07,schwarz...10}. 
The power spectra of $\sqrt[3]{\rho}{\bf u}$,
accordingly, demonstrate the Kolmogorov scaling independent of the Mach number \cite[(Kritsuk et al. 
2007; Schmidt et al. 2008; Kritsuk et al. 2009; Federrath et al. 2010; Price \& Federrath 2010)]{%
kritsuk...07,schmidt..08,kritsuk...09,federrath....10,price.10}. It seems that this result can be 
also extended to supersonic MHD turbulence, where the incompressible $4/3$-law of 
\cite[Politano \& Pouquet (1998)]{politano.98} also approximately holds in its scaling part, 
$S_{\parallel,3}^{\pm}\equiv\left<\delta {\bf Z}^{\mp}_{\parallel}(\ell) 
[\delta {\bf Z}_i^{\pm}(\ell)]^2\right>\sim\ell$ (here 
$\delta{\bf Z}_{\parallel}(\ell)\equiv[{\bf Z}({\bf x+\hat{e}}\ell)-
{\bf Z}({\bf x})]\cdot{\bf \hat{e}}$, and ${\bf\hat{e}}\ell$ is the displacement vector), 
if reformulated in terms of the mass-weighted Els\"asser fields 
${\bf Z}^{\pm}\equiv\rho^{1/3}({\bf u}\pm{\bf B}/\sqrt{4\pi\rho})$ 
\cite[(Kritsuk et al. 2009)]{kritsuk...09}. The presence of magnetic field effectively reduces
compressibility of the gas making the velocity spectra more shallow with slopes approaching the 
Iroshnikov-Kraichnan index of $-1.5$ in trans-Alfv\'enic flows. The observed spectral slope
of about $-1.8$ for the Perseus molecular cloud is thus consistent with the super-Alfv\'enic 
turbulence regime dominant in that cloud \cite[(Padoan et al. 2006)]{padoan...06}.

Recent results from numerical experiments on highly compressible turbulence stimulated theorists
to reconsider the steady-state statistics of turbulence in the inertial interval. 
\cite[Falkovich et al. (2010)]{falkovich..10} have shown that the Kolmogorov 4/5-law 
is a particular case of the general relation on the current-density correlation 
function. They derived an analog of the flux relation for compressible turbulence
that can be used as a test for direct numerical simulations and as a guide for the
development of subgrid scale models for astrophysical turbulence \cite{schmidt.10}.

Most of the recent sub-mesoscale models belong to a class of so-called converging (or colliding) 
flows of diffuse H{\sc i} originally developed to study thermal, dynamic, and gravitational
instabilities in shock-bounded slabs \cite[(Hunter et al. 1986; Vishniac 1994; Walder \& Folini 1998; 
Folini et al. 2010)]{hunter...86,vishniac94,walder.98,folini..10}. 
These models remain popular as a framework to directly simulate star formation in molecular 
clouds \cite[(V{\'a}zquez-Semadeni %
et al. 2006; Hennebelle \& Inutsuka 2006; V{\'a}zquez-Semadeni et al. 2007; Hennebelle et al. 2008; 
Inoue \& Inutsuka 2008-09; Heitsch et al. 2008-09; Banerjee et al. 2009; Niklaus et al. 2009; 
Audit \& Hennebelle 2010; 
Rosas-Guevara et al. 2010)]{hennebelle.06,inoue.08,vazquezsemadeni....06,vazquezsemadeni.....07,%
heitsch....08,hennebelle....08,heitsch..09,banerjee...09,niklaus..09,audit.10,rosasguevara...10}. Recent 
numerical experiments with converging flows have demonstrated strong sensitivity of results to 
adopted initial and boundary conditions as well as to model parameters that control the density 
of colliding gas streams, mean thermal pressure, orientation and strength of the mean magnetic 
field, levels and character of ``turbulence'' at infinity, etc. All these parameters live their
unique imprints in the statistics of derived stellar populations and any comprehensive parameter
study based on computational modeling in this framework would be prohibitively expensive.

One way to circumvent these difficulties is to exploit Prigogine's concept of self-organization in
nonequilibrium nonlinear dissipative systems \cite[(Nicolis \& Prigogine 1977)]{nicolis.77} 
in application to the ISM \cite[(e.g., Biglari \& Diamond 1989)]{biglari.89}. With this approach, 
one can use interstellar turbulence as an agent that imposes ``order'' in the form of coherent 
structures and correlations between various flow fields emerging in a simple periodic box simulation 
when a {\em statistical} steady state develops. In this case, the initial conditions are no longer 
important, instead the steady state would provide the ``correct'' turbulent initial conditions for 
star formation when self-gravity is turned on. While this idea is not new,\footnote{See, for instance, 
summary of the panel discussion on Phases of the ISM during the 1986 Grand Teton Symposium in Wyoming 
\cite{shull87}.} it remained largely undeveloped so far. In the following sections we will discuss 
this concept in more detail and report first results from a series of MHD simulations of turbulent 
multiphase ISM with the piecewise parabolic method on a local stencil 
\cite[(PPML; Ustyugov et al. 2009)]{ustyugov...09}.

\section{Self-organization in the magnetized multiphase ISM \label{self}}
In out numerical experiments, we treat the ISM as a turbulent, driven system, with kinetic energy 
being injected at the largest scales by supernova explosions, shear associated with differential 
rotation of the galactic disk, gas accretion onto the disk, etc. 
\cite[(Mac Low \& Klessen 2004; Klessen \& Hennebelle 2010)]{maclow.04,klessen.09}. This kinetic
energy is then being transferred from large to small scales in a cascade-like fashion. As our models 
include a mean magnetic field, $B_0$, some part of this kinetic energy gets stored in the turbulent
magnetic field component, $b$, generated by stretching, twisting, and folding of magnetic field
lines. The ISM is also exposed to the far-ultraviolet (FUV) background radiation due to OB 
associations of quickly evolving massive stars that form in molecular clouds. This FUV radiation
is the main source of energy input for the neutral gas phases and this volumetric thermal energy 
source is in turn balanced by radiative cooling \cite[(Wolfire et al. 2003)]{wolfire...03}.
The ISM is thus exposed to various energy fluxes, and self-organization arises as a result of 
the relaxation through nonlinear interactions of different physical constituents of the system
subject to usual MHD constraints in the form of conservation laws. In this picture, molecular 
clouds with their hierarchical internal structure form as dissipative structures that represent 
active regions of highly intermittent turbulent cascade that drain the kinetic energy supplied 
by the driving forces.

\section{Modeling the formation of molecular clouds}

To illustrate these ideas, we consider a set of simple periodic box models, which ignore gas 
stratification and differential rotation in the disk and employ an artificial large-scale 
solenoidal force to mimic the supply of kinetic energy from various galactic sources. 
This naturally leads to an upper bound on the box size, $L$, which determines our choice of 
$L=200$~pc. Our models are, thus, fully defined by the following three parameters: 
the mean gas density in the box, $n_0$; the rms velocity, $u_{\rm rms,0}$; and the mean magnetic 
field strength, $B_0$. All three would ultimately depend on $L$; Table~1 provides the 
summary of parameters for models A, B, C, D, and E assuming $L=200$~pc. The table also
gives the grid resolution, $N$, the initial values for plasma beta, 
$\beta_{\rm th,0}\equiv 8\pi p_0/B_0^2$, turbulent
beta, $\beta_{\rm turb,0}\equiv 8\pi\rho_0 u_{\rm rms,0}^2/B_0^2$, and Alfv\'enic Mach number,
$M_{A,0}=(4\pi\rho_0)^{1/2}u_{\rm rms,0}/B_0$, where $p_0$ is the initial thermal pressure of
the gas (see the phase diagram in Fig.~\ref{fig2} for more detail).

\begin{table}
  \begin{center}
  \caption{Model parameters.}
  \label{tab1}
 {\scriptsize
\begin{tabular}{lccccccc}\hline
Model& $N^3$ & $n_0$     & $u_{\rm rms,0}$ & $B_0$  & $\beta_{\rm th,0}$ & $\beta_{\rm turb,0}$ & $M_{A,0}$  \\
     & & cm$^{-3}$ & km/s             & $\mu$G &           &                       &            \\\hline
A & $512^3$ & 5         & 16             & 9.54   & 0.2       &  3.3                  & 1.3        \\
B & $512^3$ & 5         & 16             & 3.02   &  2        &  33                   & 4.0        \\
C & $512^3$ & 5         & 16             & 0.95   & 20        &  330                  & 13         \\
D & $256^3$ & 2         & 16             & 3.02   &  2        &  13.2                 & 2.6        \\
E & $256^3$ & 5         &  7             & 3.02   &  2        &  8.3                  & 2.0        \\  
\hline
\end{tabular}
}
\end{center}
\end{table}

\begin{figure}
\begin{center}
\includegraphics[width=5in]{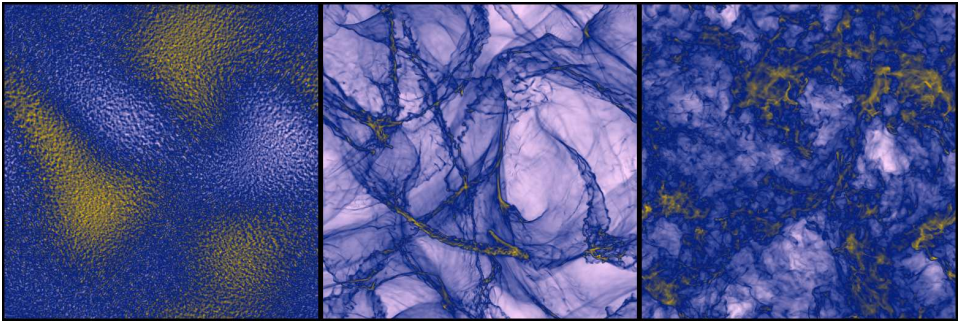} 
\end{center}
\caption{Three snapshots of projected gas density in model A taken at $t=2$, 3, and 4~Myr.
The white-blue-yellow colors correspond to low-intermediate-high projected density values.}
\label{fig1}
\end{figure}

We initiate our numerical experiments with a uniform gas distribution in the computational
domain. An addition of small random isobaric density perturbations at $t=0$ triggers a phase transition
in the thermally bi-stable gas that quickly turns $\sim25-65$\% of the gas mass into the stable 
cold phase (CNM with temperature below $T=184$~K), while the rest of the mass is shared between
the unstable and stable warm gas (WNM). In models A, B, and C, CNM and WNM each contain roughly
$\sim 50$\% of the total H{\sc i} mass in agreement with observations \cite[(Heiles \& Crutcher 2005)]{heiles.05}. 
We then turn on the forcing and after a few large-eddy turn-over times the simulation approaches 
a statistical steady state. If we replace this two-stage initiation process with a one-stage 
procedure by turning the driving on at $t=0$, the properties of the steady state remain unchanged.
Figure~\ref{fig1} illustrates this evolutionary sequence for the two-stage case with three snapshots
of projected gas density for model A. The left panel shows two-phase medium at $t=2$~Myr right before
we turn on the forcing; the panel in the middle illustrates an early stage of turbulization with transient
``colliding flows'' at $t=3$~Myr. The right panel shows the projected density at $t=4$~Myr for a
statistically developed turbulent state. Molecular clouds can be seen in the right panel as 
filamentary brown-to-yellow structures (note that these are morphologically quite 
different from the transient dense structures in the middle panel). The rms magnetic field is amplified by
the forcing and saturates when the relaxation in the system results in a steady state, 
see Fig.~\ref{fig2}. The level of saturation depends on $B_0$ and on the rate of kinetic 
energy injection by the large-scale force, which is in turn determined by $u_{\rm rms}$ 
and $n_0$. This level can be easily controlled with the model parameters. In the 
saturated regime, models A and B tend to establish energy equipartition ($E_{\rm K}\sim E_{\rm M}$), 
while the saturation level of magnetic energy in model C is a factor of $\sim3$ lower than 
the equipartition level. The mean thermal energy also gets a slight boost due to forcing, 
but remains subdominant in all the models. A typical phase diagram is shown in Fig.~\ref{fig2}. 
The contours indicate constant levels of volume fraction for different regimes of the thermal 
pressure, $p_{\rm th}$, and density, $n$, separated by factors of 2. About 23\% of the domain 
volume is filled with the stable warm phase at $T>5250$~K, the stable cold phase ($T<184$~K) 
occupies $\sim7$\%, and $\sim70$\% of the volume resides in the thermally unstable regime at intermediate
temperatures. The big orange dot at the center indicates the (forgotten) initial conditions for
models A, B, C, and E. The phase diagram indicates that turbulence supports an enormously wide
range of thermal pressures and also that $p_{\rm th}$ in the molecular gas ($n>100$~cm$^{-3}$)
is higher than that in the diffuse ISM, even though self-gravity is ignored in the model.

\begin{figure}
\begin{center}
\includegraphics[width=2.5in]{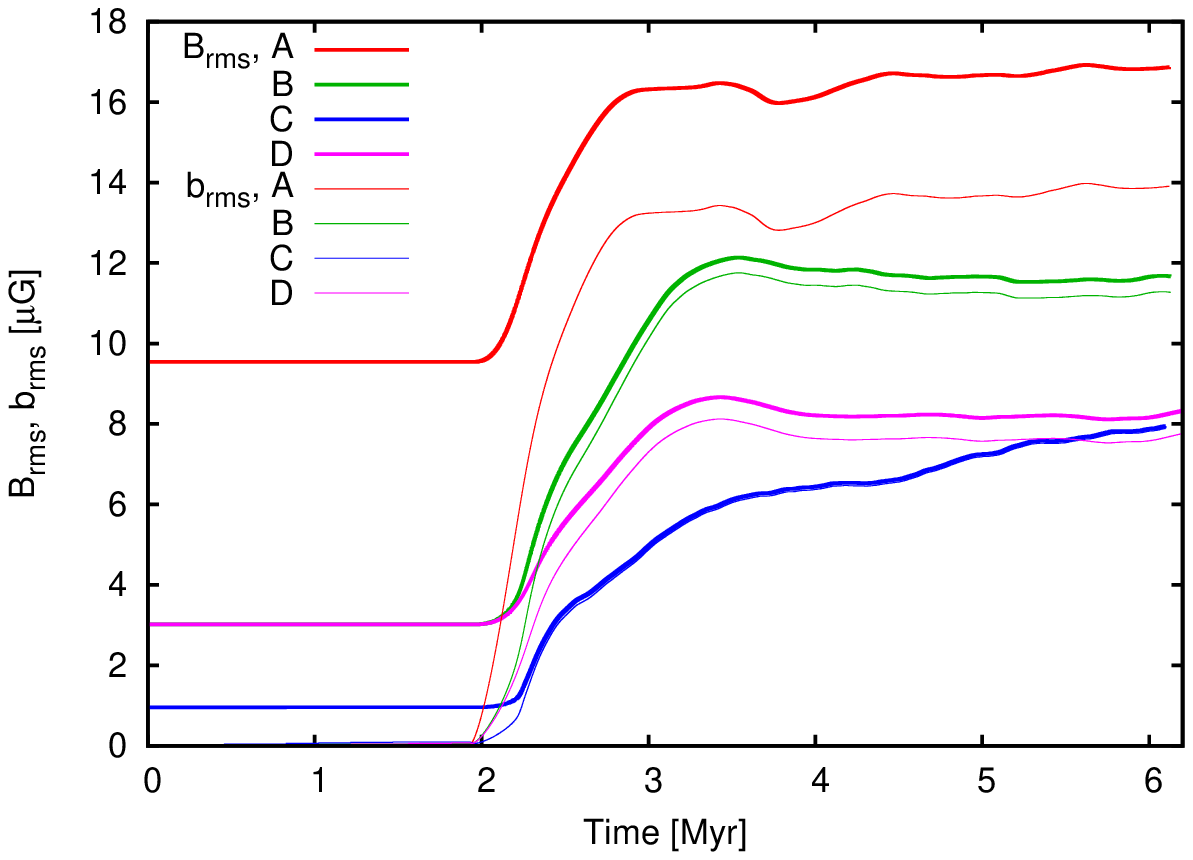}
\includegraphics[width=2.5in]{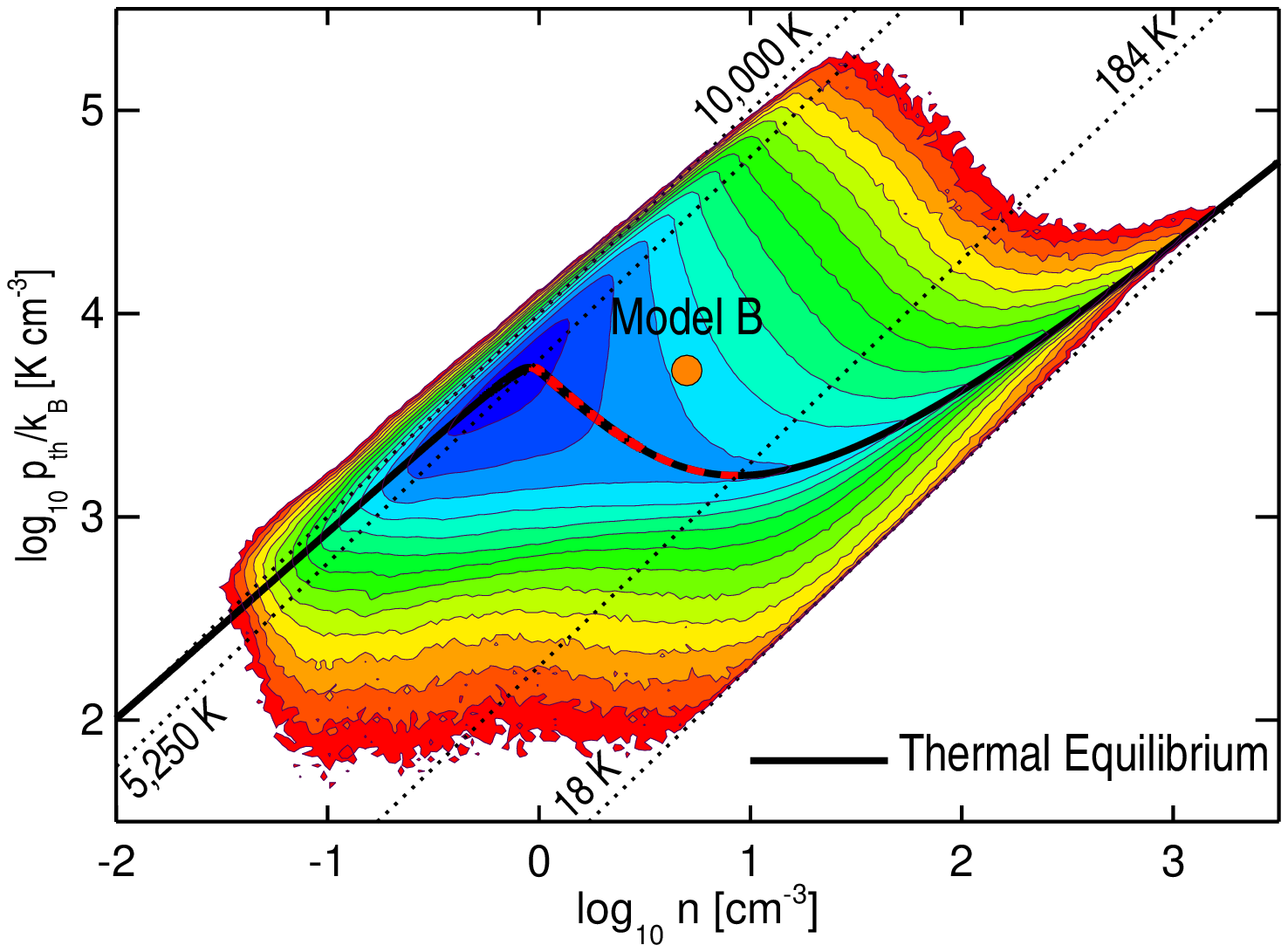}
\end{center}
\caption{Time evolution of the rms magnetic field strength, $B_{\rm rms}$, and its turbulent 
component, $b_{\rm rms}$, for models A, B, C, and D ({\em left} panel). Phase diagram (thermal pressure
versus gas density) for model B at $t=5$~Myr ({\em right} panel).}
   \label{fig2}
\end{figure}

\begin{figure}
\begin{center}
\includegraphics[width=2.5in]{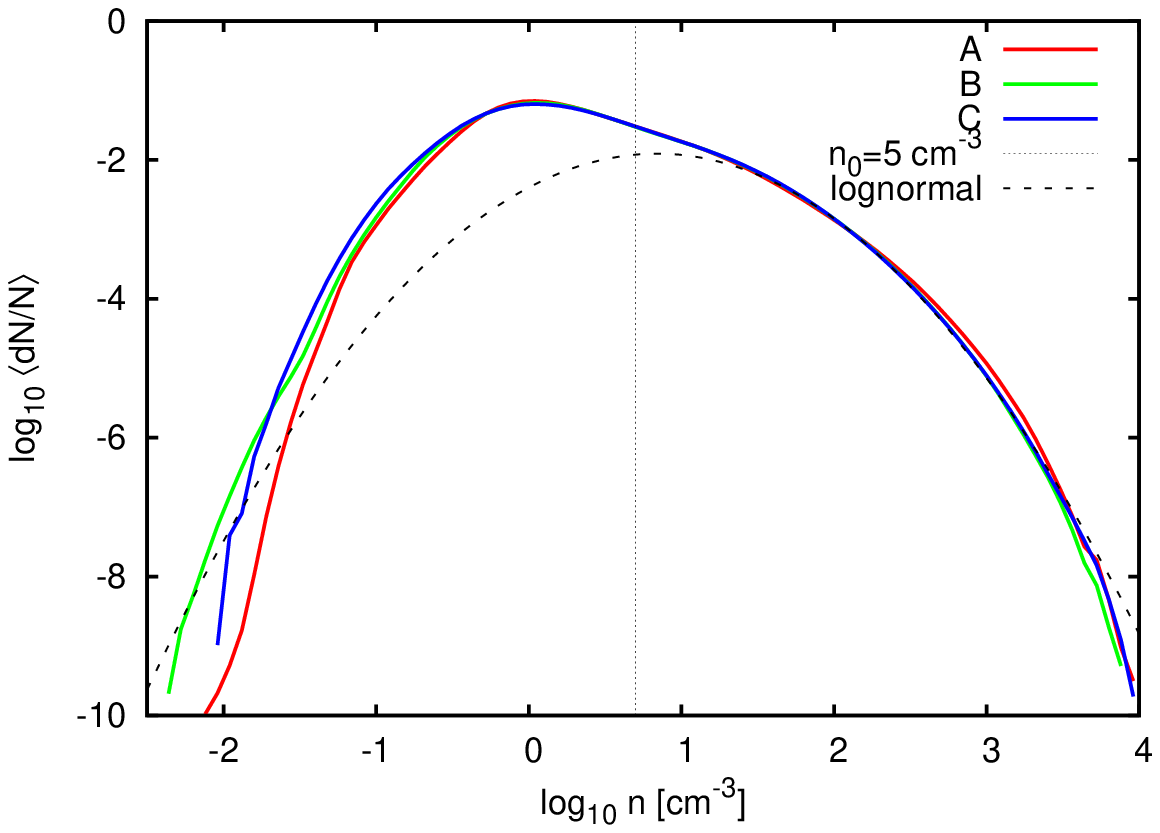}
\includegraphics[width=2.5in]{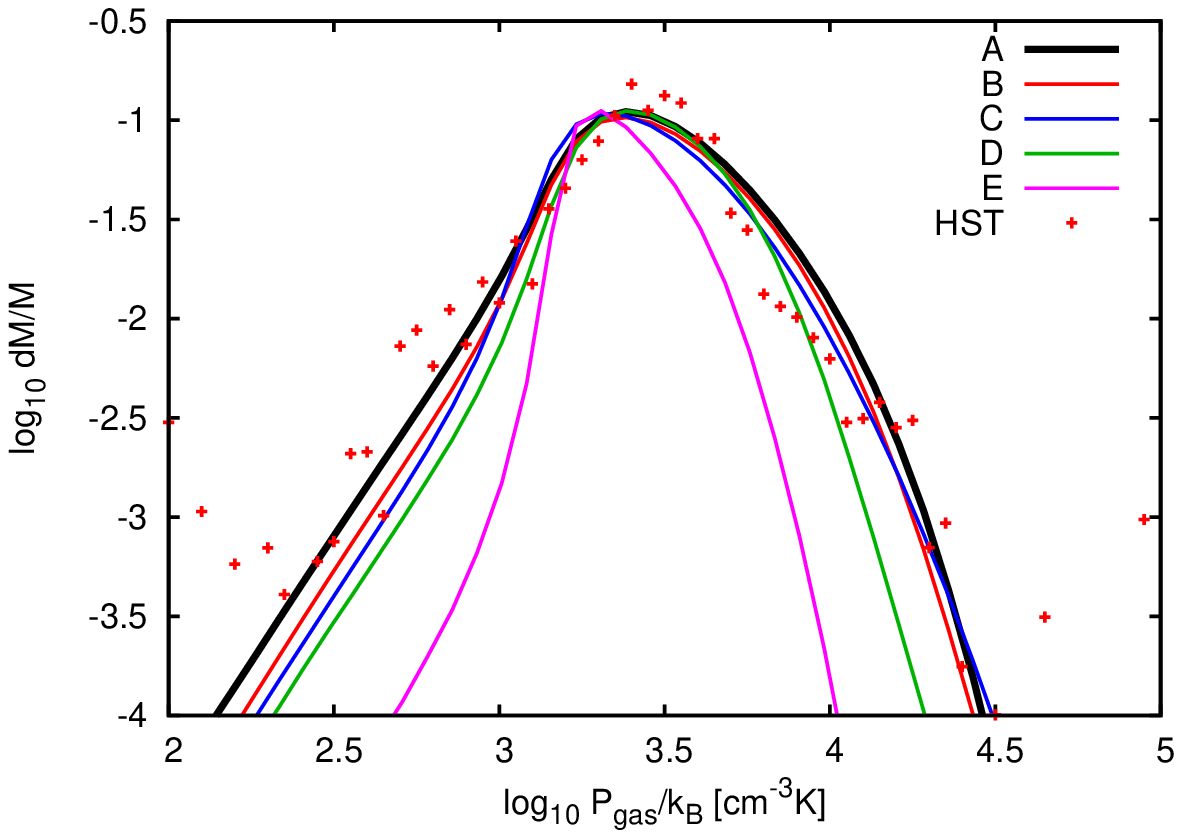}
\end{center}
 \caption{Time-average density distributions for fully developed turbulence in models 
A, B, and C ({\em left} panel) and time-average mass-weighted thermal pressure distributions 
for models A, B, C, and D ({\em right} panel); see text for more detail.}
   \label{fig3}
\end{figure}

Figure~\ref{fig3} ({\em left}) shows the time-average density PDFs for models A, B, and C in the steady
state. The effect of magnetic field on the density PDF is apparently very weak on average,
\begin{figure}
\begin{center}
\includegraphics[width=2.5in]{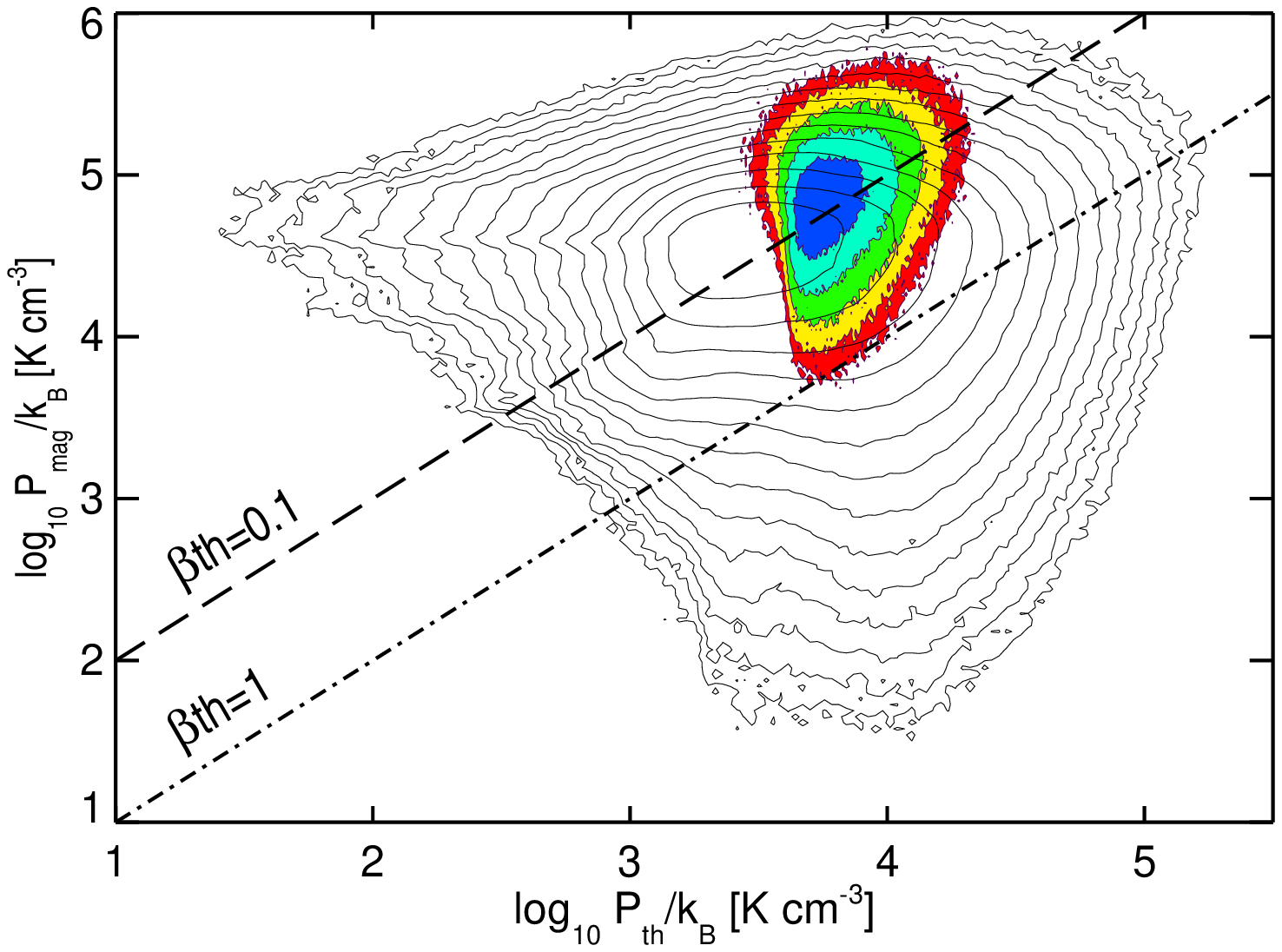}
\includegraphics[width=2.5in]{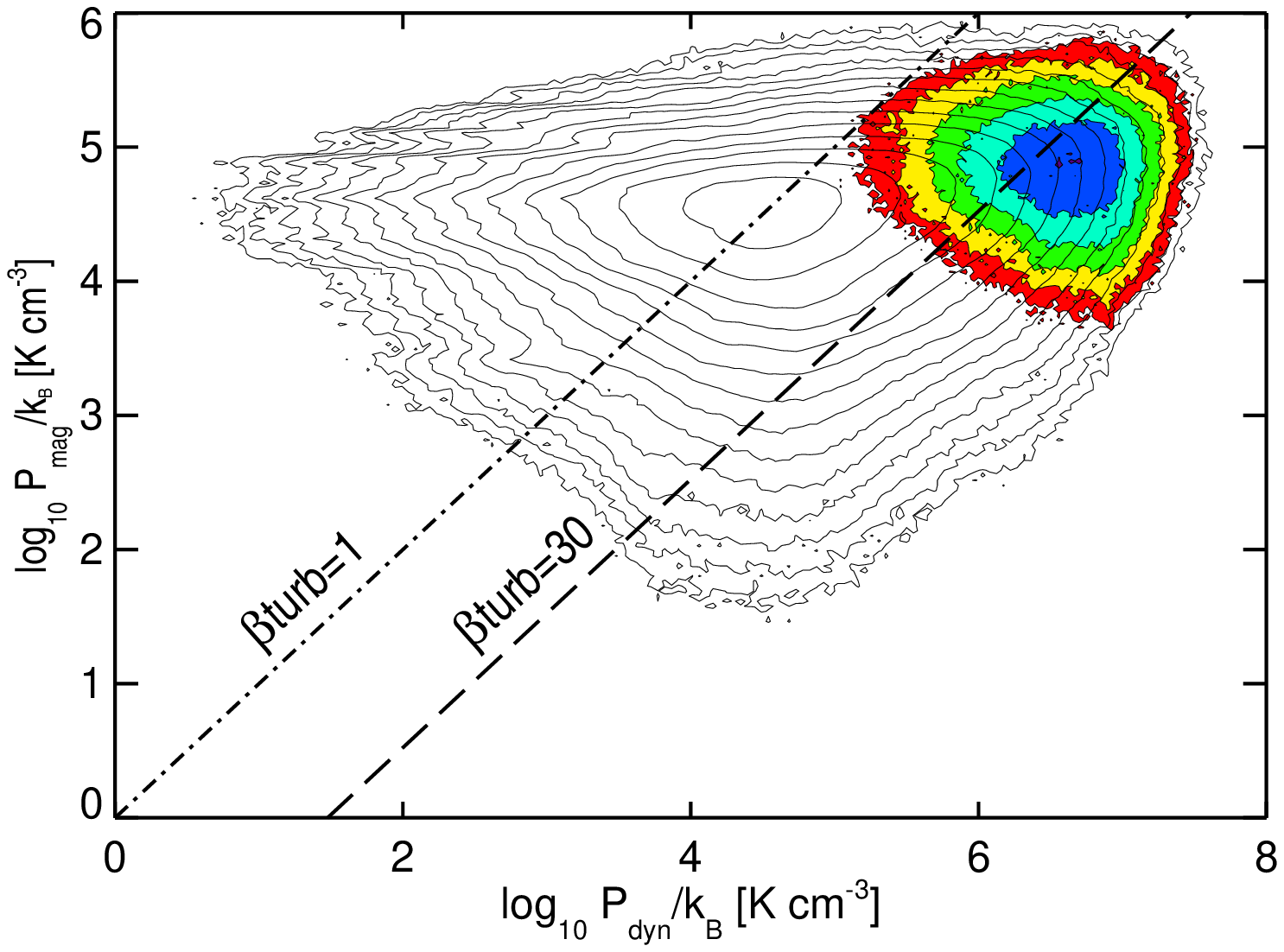}
\end{center}
 \caption{Distributions of magnetic pressure vs. thermal ({\em left}) and dynamic ({\em right})
pressure for a snapshot from model B at $t=5$~Myr.}
   \label{fig4}
\end{figure}
\begin{figure}
\begin{center}
\includegraphics[width=2.5in]{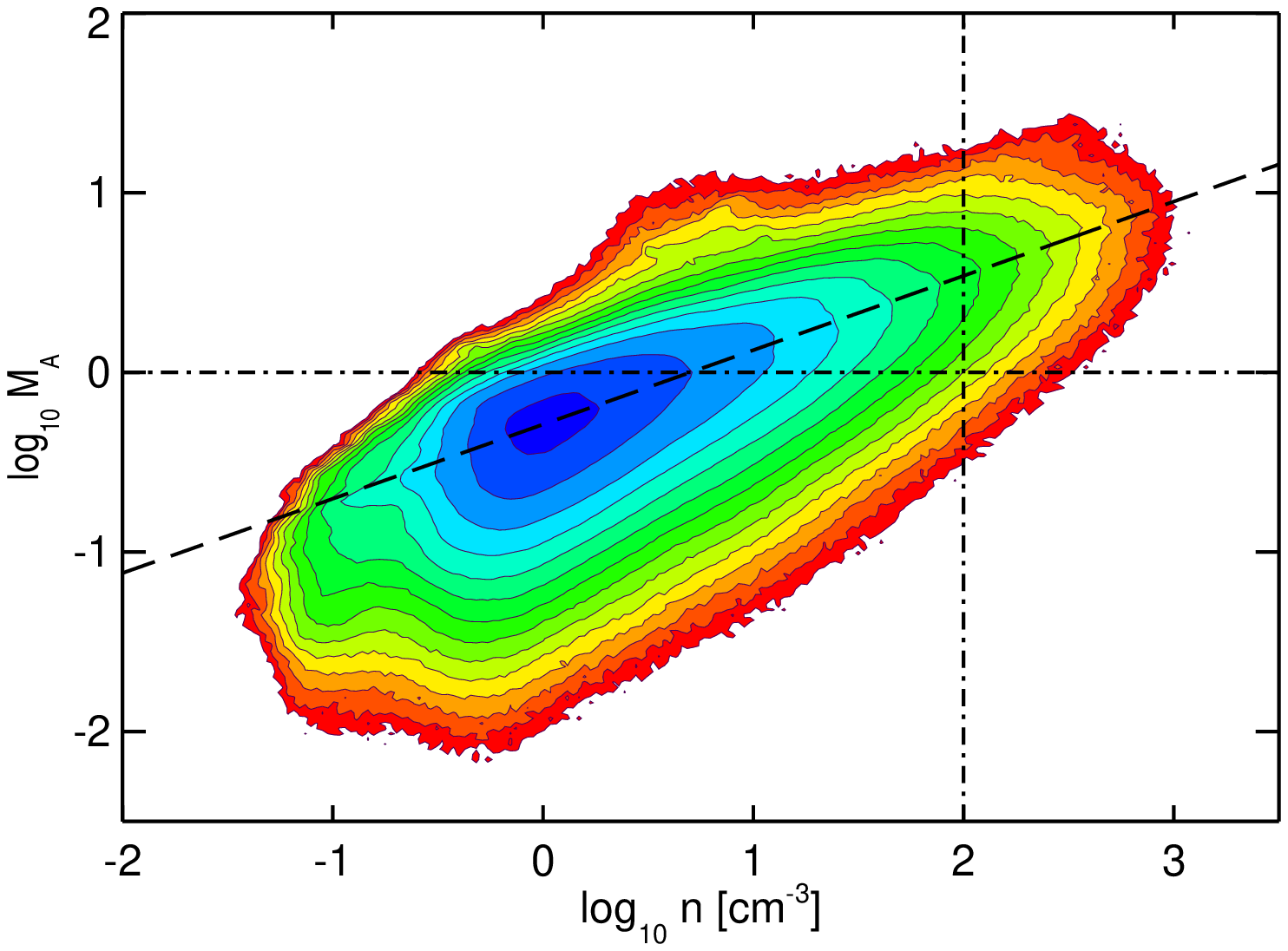}
\includegraphics[width=2.5in]{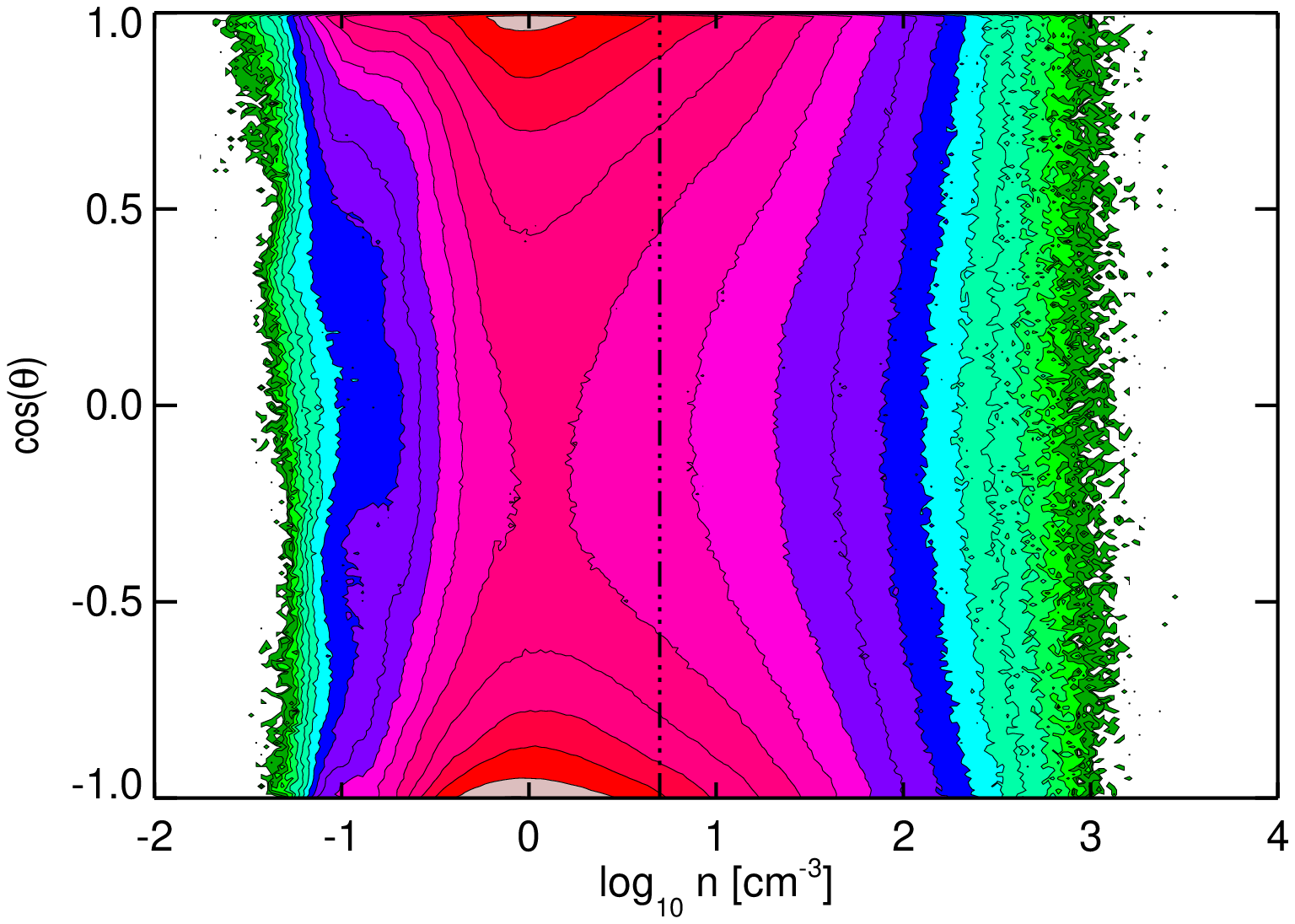}
\end{center}
\caption{Distributions of the Alfv\'enic Mach number and the cosine of the alignment angle 
vs. density for a snapshot from model A at $t=5$~Myr.}
\label{fig5}
\end{figure}
and the high-density part of the PDF can be well approximated by a lognormal as in the 
isothermal case. The signature of the two stable thermal phases in the PDF is smeared by the
(relatively) high level of turbulence, $u_{\rm rms}(100$~pc$)=16$~km/s (Brunt \& Heyer 2004), 
but the overall shape of the distribution
is not lognormal. The distribution of thermal pressure for models A, B, and C spans about 6 dex leaving
no room for the old pressure-supported cloud picture in the violent ISM. All distributions match the 
characteristic pressure typical for the Milky Way disk at the solar radius and show only weak
dependence on $B_0$, while the width of the distribution remains sensitive to $u_{\rm rms}$ and $n_0$.
Figure~\ref{fig3} ({\em right}) shows how the mass-weighted pressure distributions obtained in our
models compare with the distribution reconstructed from high-resolution UV spectra of hot stars in
the HST archive \cite{jenkins.10}. It seems that models with $u_{\rm rms}=16$~km/s reproduce both 
the shape and the width of the observed distribution quite nicely, while a lower turbulence level in
model E makes that model distribution too narrow.

These numerical experiments allow us to probe the levels of magnetic field strength in molecular
clouds that form self-consistently in the magnetized turbulent diffuse ISM. In the {\em left}
panel of Fig.~\ref{fig4}, we show a scatter plot of magnetic vs. thermal pressure for model B 
at $t=5$~Myr. The black contour lines show the distribution for the whole domain, which is 
centered at $\beta_{\rm th}\approx0.1$. The subset of cells representing the molecular
gas ($T<100$~K and $n>100$~cm$^{-3}$, color contour plot) shows very similar mean values of 
$\beta_{\rm th}$. This indicates a target plasma beta for realistic isothermal molecular cloud 
turbulence simulations.
The {\em right} panel of the same Figure shows a scatter plot of magnetic vs. dynamic pressure
for the same snapshot from the same model. The distribution for the whole domain is 
centered at $\beta_{\rm turb}\approx1$ because kinetic and magnetic energy levels are 
close to equipartition on average. At the same time, the subset of cells representing 
the molecular gas (color contours) shows a distribution centered at $\beta_{\rm turb}\approx30$
meaning that turbulence in the molecular gas is super-Alfv\'enic. Figure~\ref{fig5}, {\em left}
panel, shows the distribution of Alfv\'enic Mach number, $M_{\rm A}$, as a function of
density for the strongly magnetized model A that further supports this result. There is 
a clear positive correlation, $M_{\rm A}\sim n^{0.4}$, indicated by the dashed line and most 
of the dense material ($n>100$~cm$^{-3}$) clearly falls into the super-Alfv\'enic part of
the distribution.

A key to understanding the origin of this super-Alfv\'enic regime in the cold and dense 
molecular gas lies in the process of self-organization in magnetized ISM turbulence
that we briefly introduced in Section~\ref{self}. The statistical steady state that
our models attain on a time-scale of a few million years is characterized by a certain
degree of alignment between the velocity and magnetic field lines. Figure~\ref{fig5}, 
{\em right} panel, shows the distribution of the cosine of the alignment angle, 
$\cos\theta\equiv{\bf B\cdot u}/(Bu)$, for model A at $t=5$~Myr. The contours indicate
a saddle-like structure of this probability distribution with a strong alignment
regime ($\cos\theta=\pm1$) in the WNM at and around $n\sim1$~cm$^{-3}$. This means 
that compressions in the WNM gas, which is on average trans-Alfv\'enic
(e.g., $M_{\rm A}\in[0.6, 0.9]$ in model A), occur preferentially along the field 
lines. If molecular clouds form in the turbulent ISM via large-scale compression 
of the diffuse H{\sc i}, then turbulence in such molecular clouds can only be 
super-Alfv\'enic, see also \cite{padoan......10}.

\section{Conclusion}
Rapid development of computational astrophysics in the recent years enabled
progress in understanding the basics of interstellar turbulence. These new advances
will help us to move forward with direct star formation simulations from turbulent
initial conditions.

\vspace{0.3cm}
{\underline{\it Acknowledgements}}. 
This research was supported in part by the National Science Foundation through 
grants AST-0607675, AST-0808184, and AST-0908740, as well as through TeraGrid 
resources provided by NICS and SDSC (MCA07S014) and through DOE Office of 
Science INCITE-2009 and DD-2010 awards allocated at NCCS (ast015/ast021).

\end{document}